\documentclass[reprint,superscriptaddress,amsmath,amssymb,aps,prx]{revtex4-2}
\usepackage{graphicx}
\usepackage{dcolumn}
\usepackage{bm}
\usepackage{qcircuit}
\usepackage{algpseudocode}
\usepackage{graphicx}
\usepackage{caption}
\usepackage{algorithm}
\usepackage{subfigure}
\usepackage{url}
\usepackage[pass]{geometry}
\usepackage{float}
\usepackage{amsthm}
\usepackage{physics}
\usepackage[version=4]{mhchem}
\usepackage{diagbox}
\usepackage{xcolor}
\usepackage{footnote}
\usepackage{enumitem}
\makeatletter
\renewcommand*\env@matrix[1][\arraystretch]{%
\edef\arraystretch{#1}%
\hskip -\arraycolsep
\let\@ifnextchar\new@ifnextchar
\array{*\c@MaxMatrixCols c}}
\makeatother
\begin{document}
\preprint{APS/123-QED}
\title{Locating Rydberg Decay Error in SWAP-Leakage Reduction Circuit
    protocol}
\author{Cheng-Cheng Yu}
\affiliation{Hefei National Research Center for Physical Sciences at the Microscale and School of Physical Sciences, University of Science and Technology of China, Hefei 230026, China}
\affiliation{Shanghai Research Center for Quantum Science and CAS Center for Excellence in Quantum Information and Quantum Physics,
University of Science and Technology of China, Shanghai 201315, China}
\affiliation{Hefei National Laboratory, University of Science and Technology of China, Hefei 230088, China}
\author{Yu-Hao Deng}
\email{dengyh@ustc.edu.cn}
\affiliation{Hefei National Research Center for Physical Sciences at the Microscale and School of Physical Sciences, University of Science and Technology of China, Hefei 230026, China}
\affiliation{Shanghai Research Center for Quantum Science and CAS Center for Excellence in Quantum Information and Quantum Physics,
University of Science and Technology of China, Shanghai 201315, China}
\affiliation{Hefei National Laboratory, University of Science and Technology of China, Hefei 230088, China}
\author{Chao-Yang Lu}
\affiliation{Hefei National Research Center for Physical Sciences at the Microscale and School of Physical Sciences, University of Science and Technology of China, Hefei 230026, China}
\affiliation{Shanghai Research Center for Quantum Science and CAS Center for Excellence in Quantum Information and Quantum Physics,
University of Science and Technology of China, Shanghai 201315, China}
\affiliation{Hefei National Laboratory, University of Science and Technology of China, Hefei 230088, China}
\affiliation{New Conerstone Science Laboratory, Hefei, 230026, China}
\author{Ming-Cheng Chen}
\email{cmc@ustc.edu.cn}
\affiliation{Hefei National Research Center for Physical Sciences at the Microscale and School of Physical Sciences, University of Science and Technology of China, Hefei 230026, China}
\affiliation{Shanghai Research Center for Quantum Science and CAS Center for Excellence in Quantum Information and Quantum Physics,
University of Science and Technology of China, Shanghai 201315, China}
\affiliation{Hefei National Laboratory, University of Science and Technology of China, Hefei 230088, China}
\author{Jian-Wei Pan}
\affiliation{Hefei National Research Center for Physical Sciences at the Microscale and School of Physical Sciences, University of Science and Technology of China, Hefei 230026, China}
\affiliation{Shanghai Research Center for Quantum Science and CAS Center for Excellence in Quantum Information and Quantum Physics,
University of Science and Technology of China, Shanghai 201315, China}
\affiliation{Hefei National Laboratory, University of Science and Technology of China, Hefei 230088, China}
\begin{abstract}
Qubit leakage and loss, particularly Rydberg-induced decay during two-qubit gates, pose significant challenges to fault-tolerant quantum computing with neutral atom arrays, as they propagate to correlated errors and degrade code distance. Here, we present a hardware-efficient scheme for addressing Rydberg decay using the SWAP-Leakage Reduction Circuit (SWAP-LRC) protocol, which leverages ancilla-data qubit swaps for in-line leakage mitigation. This strategy eliminates the need for atom-species-specific mid-circuit detection or additional ancillary qubits. Based on experimental detection capabilities, we present two specialized decoders. For detectable leakage/loss (e.g., in $^{171}$Yb), our Located Decoder achieves a high threshold of 2.33\% per CNOT gate and an improved error distance, significantly outperforming conventional Pauli error models. More interestingly, for scenarios where only one error type is detectable (e.g., atom loss for $^{87}$Rb), our Critical Decoder specifically targets and mitigates the most detrimental critical faults caused by correlated leakage, achieving an error distance comparable to standard Pauli errors. Our findings offer insights for handling complex non-Pauli errors for neutral atom quantum computation.
\end{abstract}
\maketitle
\section{Introduction}\label{sec1}
Neutral atom arrays have emerged as a promising platform for quantum computation \cite{revmodphys.82.2313,morgado2021quantum,kaufman2021quantum,wu2021concise}. Leakage error from the Rydberg states (Rydberg decay) is an inherent and major error source for neutral atoms \cite{cong2022hardware,PRXQuantum.5.040343,msbj-fxw7,wu2022erasure,jandura2024surfacecodestabilizermeasurements}. Without proper methods to solve it, the leakage error from Rydberg decay degrades the error distance from $d_e = \lfloor \frac{d+1}{2} \rfloor$ to $d_e = \lfloor \frac{d+3}{4} \rfloor$, greatly reducing the effectiveness of error correction \cite{PhysRevA.88.042308,suchara2015leakage,brown2020critical,jandura2024surfacecodestabilizermeasurements}. Different methods have been proposed for this issue. Erasure conversion, a hardware-specific protocol, is designed for alkaline-earth atoms, which utilizes mid-circuit leakage detection to convert leakage error to benign erasure error after each multi-qubit gate \cite{wu2022erasure,kang2023quantum,ma2023high,scholl2023erasure,sahay2023high,omanakuttan2024coherencepreservingleakagedetection}. Several experiments have demonstrated the erasure conversion protocol on $\ce{^{171}_{}Yb}$ atoms and $\ce{^{88}_{}Sr}$ atoms \cite{ma2023high,scholl2023erasure}. The circuit-based protocol, which attaches a small-scale circuit to the original one, detects the leakage and renews the atom at the same time \cite{suchara2015leakage,PRXQuantum.5.040343,Perrin_2025}. Both theoretical protocols enhance the performance of error correction, achieving a high threshold and an error distance $d_e = d$, for pure Rydberg decay errors.\\

The two approaches are either applicable to a certain type of atoms or introduce additional hardware overhead. Erasure conversion, proposed for $\ce{^{171}_{}Yb}$ atoms, requires mid-circuit leakage detection \cite{wu2022erasure} and atom replenishment \cite{sahay2023high}, which makes its near-term implementation challenging. The circuit-based protocol requires additional ancilla qubits for leakage reduction equal to the number of the data qubits, resulting in substantial qubit overhead \cite{Perrin_2025}. An alternative but more hardware-efficient solution is the SWAP-Leakage Reduction Circuit (SWAP-LRC \cite{suchara2015leakage,brown2020critical,PhysRevA.100.032325}, sometimes also referred as Quick-LRC \cite{suchara2015leakage}), a protocol that swaps the role of data qubits and syndrome ancilla qubits during each round of syndrome measurement \cite{suchara2015leakage,brown2020critical,McEwen_2023,PhysRevA.88.042308}. This method utilizes inherent ancilla qubits for syndrome measurement to remove leakage, thereby avoiding any additional qubit overhead. A similar method has been used in error correction experiments with superconducting qubits \cite{eickbusch2025demonstration}. However, even if SWAP-LRC has the lowest overhead in circuit-level leakage reduction and is friendly to experimental realization, it also has severe error propagation \cite{suchara2015leakage}, which may lead to poor performance and degraded distance \cite{brown2020critical, PhysRevA.88.042308}.\\ 

In this article, we address Rydberg decay with SWAP-LRC and employ final leakage detection to locate the propagated error. From theoretical analysis, we reveal that \textit{correlated leakage} reduces the error distance and hinders the sub-threshold scaling. We then propose two decoders based on different experimental detection capabilities: \begin{itemize}[leftmargin=1.5em, itemsep=0.1em, topsep=0.2em]
    \item Condition I -- both types of Rydberg decay (atom loss from BBR and decay error) can be detected;
    \item Condition II -- only one type (atom loss or decay error) can be detected.
\end{itemize} For Condition I, we propose the \textit{located decoder} that achieves a high threshold 2.33(3)\% for each CNOT gate and an improved error distance for pure Rydberg decay errors \cite{wu2022erasure,ma2023high}. We also demonstrate the advantage of sub-threshold scaling and the logical error rate when Rydberg decay is the major error source. For Condition II, we propose \textit{critical decoder} inspired by a previous study on the critical fault \cite{brown2020critical}. We reveal that detecting one type of Rydberg decay is enough to preserve the error distance $d_e = \frac{d+1}{2}$ by locating the two-qubit error chain simultaneously. Therefore, critical decoder is an experiment-friendly decoding strategy to mitigate the damaging effect of Rydberg decay on the sub-threshold scaling.

The rest of the article is organized as follows. In Sec.\ref{sec2}, we provide theoretical analysis on the feature of Rydberg decay (Sec.\ref{ssec2-1}), the Rydberg decay error propagation (Sec.\ref{ssec2-2}), and the property of located error (Sec.\ref{ssec2-3}). In Sec.\ref{sec3}, we first provide results for Condition I, when both types of leakage can be detected in Sec.\ref{ssec3-1}. We demonstrate high error thresholds and improved code distances under pure Rydberg decay errors. We also validate that correlated leakage significantly degrades sub-threshold scaling. When Rydberg decay dominates, we show that using a located decoder offers clear performance advantages. Then we reveal that \textit{critical decoder} preserves the error distance $d_e = \frac{d+1}{2}$ for Condition II, in Sec.\ref{ssec3-2}. Finally, we provide some discussion and outlook in Sec.\ref{sec4}.

\section{Theoretical Analysis}\label{sec2}
\subsection{Analysis of Rydberg decay}\label{ssec2-1}

Our analysis of Rydberg decay is partially based on one previous work \cite{msbj-fxw7}. One feature of Rydberg decay is that it can be modeled as a leaked state $\ket{L}$ that is not involved in subsequent single- and two-qubit gates, no matter if the atom is lost from anti-trapping of atoms in the Rydberg state \cite{PRXQuantum.5.040343,cong2022hardware} or is leaked onto a state that is energetically separated from the qubit subspace \cite{wu2022erasure,PRXQuantum.5.040343}. With Pauli twirling and randomized compiling \cite{Wallman_2016}, a leakage error can be considered as two kinds of jump operators with equal probability, and we derive the \textit{Pauli-like} error propagation by forward propagation of the jump operator \cite{msbj-fxw7}.\\

The state-generation circuit discussed in the previous work is a simplified one, since it only includes CZ gates and the jump operator keeps its form after forward propagation \cite{jayashankar2022achieving}. When considering the syndrome measurement circuit for the toric code with SWAP-LRC, the form of the jump operator changes, thereby breaking the pattern of \textit{Pauli-like} error propagation. A detailed derivation of the error propagation property is in Appendix \ref{appendix1}, and the conclusion is that: When the circuit is \textit{noise-structure preserving} for the jump operator \cite{jayashankar2022achieving}, the error propagation pattern is \textit{Pauli-like}, depending on the jump operator; when the circuit is not \textit{noise-structure preserving}, the error propagation degrades to tailored-Pauli propagation \cite{PhysRevA.100.032325,d1v7-nctj}. However, tailored-Pauli propagation is not as harmful as the depolarization leakage model, since it does not degrade the error distance for the toric code and its variants \cite{PhysRevA.100.032325}.\\

Another characteristic of the Rydberg decay error is the \textit{correlated leakage}. When one of the atoms decays from a Rydberg state to a low-lying state, the blockade effect fails. Consequently, a neighboring atom may be excited to the Rydberg state with $O(1)$ probability and subsequently lost \cite{wu2022erasure}. It suggests that data qubit and ancilla qubit encounter two leakages simultaneously with probability $O(p_e)$ ($p_e$ is the Rydberg decay error rate of a single two-qubit gate), when we treat decay error to lower levels and atom loss uniformly \cite{wu2022erasure}. We avoided this question in our previous work because of the limited error propagation in the state-generation circuit \cite{msbj-fxw7}. However, such an error degrades the error distance when it happens at one certain error site \cite{brown2020critical}, as discussed in Sec.\ref{ssec2-2}. \\

\begin{figure}[h]
    \centering
    \includegraphics[width=0.48\textwidth]{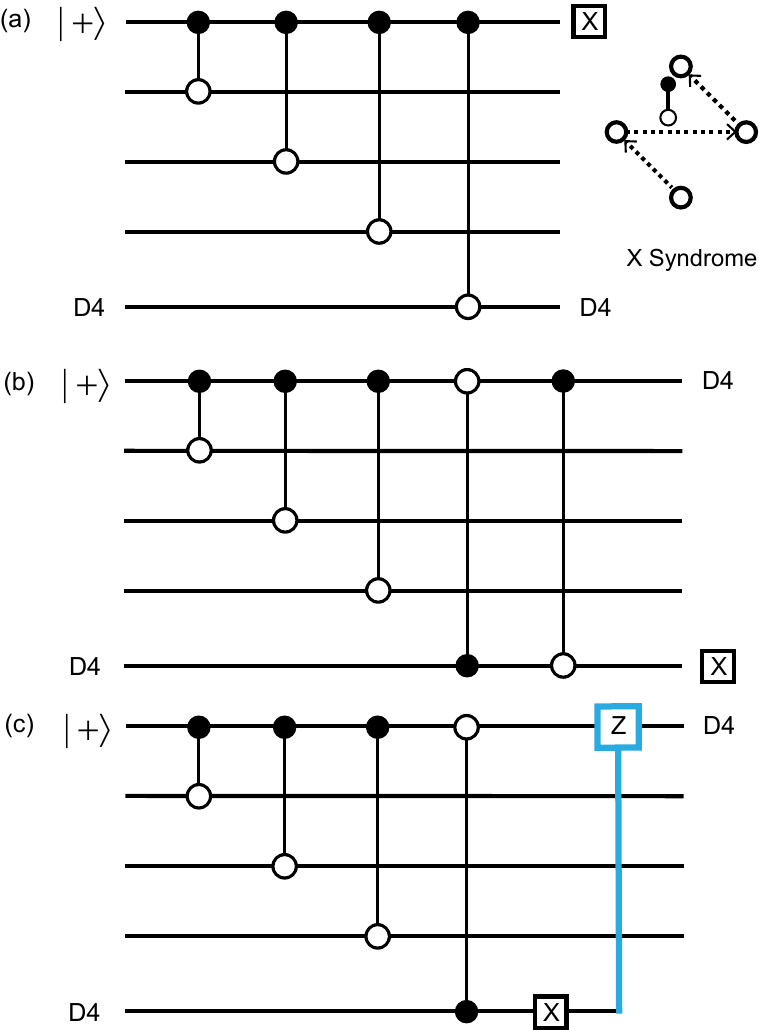}
    \captionsetup{justification = raggedright,singlelinecheck = false}
    \caption{\textbf{(a)} The standard syndrome measurement circuit for X stabilizer in the toric code. \textbf{(b)} The X syndrome measurement with SWAP-LRC. The ancilla qubit takes the role of the fourth interaction data qubit while the fourth data qubit is measured to extract the result of the syndrome. \textbf{(c)} The final CNOT gate can be replaced by a feed-forward gate. Here, the feed-forward gate is virtually implemented with software correction \cite{Baranes_2026}.}
    \label{fig1}
\end{figure}

\subsection{Rydberg decay error propagation in SWAP-LRC}\label{ssec2-2}

In this section, we discuss the error propagation in SWAP-LRC. We select the toric code to study the performance because of the convenience of realizing SWAP-LRC with the period boundary condition. The method can be extended to its planar variants, surface code, by appending a line of ancilla qubits \cite{suchara2015leakage,PhysRevA.100.032325}. In the syndrome measurement circuit with SWAP-LRC, the role of the data qubit and the ancilla qubit is exchanged before each round of measurement, see Fig.\ref{fig1}. In a standard syndrome measurement circuit, the data qubit is always unmeasured, so the leaked data qubits are not removed and spread additional errors continuously. With SWAP-LRC, the data leakage is detected and removed in each round of measurement, and the ancilla leakage is removed in the next round of measurement, when the previous ancilla qubit acts as a data qubit. Therefore, both the data and ancilla leakage are removed within two rounds. Final CNOT gate can be replaced by a feed-forward gate, which is implemented virtually by software correction \cite{Baranes_2026,Perrin_2025}. This replacement makes no difference to the error propagation, but we don't need to account for gate errors in the final CNOT gate of each syndrome measurement. Equipped with the circuit in Fig.\ref{fig1}(c), no additional qubits or physical gates are required to remove leakage and renew the atoms. \\

\begin{figure}[h]
    \centering
    \includegraphics[width=0.45\textwidth]{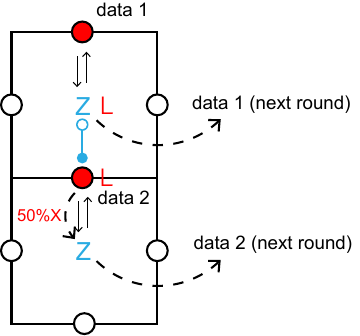}
    \captionsetup{justification = raggedright,singlelinecheck = false}
    \caption{The \textit{correlated leakage} instance in the first CNOT gate degrades distance: The leaked ancilla qubit becomes a leaked data qubit in the next round, and the leaked data qubit propagates to a $50\%$ X error to the ancilla qubit in this round, namely the data qubit in the next round. These two data qubits lie on the same logical operator, so it degrades the distance.}
    \label{fig2}
\end{figure}

We provide a detailed analysis of the error propagation in Appendix \ref{appendix2}. Based on that, we identify that the \textit{correlated leakage} in the first CNOT gate of Z syndrome measurement degrades the error distance for a vertical logical X operator. As illustrated in Fig. \ref{fig2}, if a \textit{correlated leakage} occurs in the first CNOT gate, the ancilla leakage becomes a data leakage in the next round, on the qubit that exchanges with the ancilla (data 1 in Fig.\ref{fig2}) and the data leakage propagates to 50\% X error to the ancilla qubit that exchanges with it, namely a 50\% X error on that data qubit (data 2 in Fig.\ref{fig2}) in the next round. These two data qubit errors lie in the same logical X operator, so the correlated leakage degrades the distance. This discussion also applies to the vertical logical Z operator when considering logical Z error. So this serves as a \textit{critical fault} that threatens the sub-threshold scaling of error correction \cite{brown2020critical}.

\subsection{Error distance, located error and re-weighting}\label{ssec2-3}

The error distance is the minimum number of physical (gate) errors needed to generate a logical error. It is related to the sub-threshold scaling of the logical error rate. When the physical error rate is small enough, the logical error rate is suppressed exponentially with error distance $d_e$, namely $p_L\sim (p/p_{th})^{d_e}$ \cite{PhysRevResearch.7.033074,PhysRevA.86.032324}. The error distance is influenced by two factors: One is whether the error is located, and the other is whether a single gate error induces a two-qubit error chain along the logical operator. The second factor is closely related to gate sequence and the form of error propagation \cite{Tomita_2014,PhysRevResearch.7.033074, PhysRevA.88.042308,brown2020critical} and we have discussed the second factor for the Rydberg decay error in Sec.\ref{ssec2-2}. For the first factor, an error correction code with code distance $d$ corrects $d-1$ located errors (erasure errors) but $\lfloor \frac{d}{2} \rfloor$ Pauli errors \cite{wu2022erasure, Nielsen_Chuang_2010}. `Located' means that we know which qubits (or gates for detector error model \cite{Gidney_2021}) are suspicious to be faulty. Therefore, if a Rydberg decay error is detected and the atom is renewed, we regard this as an erasure error \cite{wu2022erasure}. Recently several works are showing that the general located error also has an error distance $d$ \cite{d1v7-nctj,msbj-fxw7,Perrin_2025,Baranes_2026}. The general located error corresponds to the conditions that use imperfect erasure check \cite{d1v7-nctj} or final erasure check (instead of erasure check after each gate) physically \cite{msbj-fxw7,Perrin_2025,Baranes_2026}. This insight inspires more hardware-efficient protocols regarding leakage error since additional erasure checks come along with additional hardware operation \cite{wu2022erasure} or qubit overhead \cite{cong2022hardware,985g-58gd}.\\

Previous studies have analyzed error correction performance for a mixture of the Pauli error ($d_e = \frac{d+1}{2}$) and the erasure error ($d_e = d$), in both thresholds and the sub-threshold scaling (effective error distance) \cite{wu2022erasure,sahay2023high}. When the amount of error is comparable, the logical error rate is always governed by the more harmful error. As is revealed in the results \cite{wu2022erasure}, a small ratio of Pauli error (with $d_e = \frac{d+1}{2}$) decreases the effective error distance drastically. Similarly, when a more harmful error with error distance $d_e = \lfloor\frac{d+3}{4}\rfloor$ exists \cite{jandura2024surfacecodestabilizermeasurements} if it is not properly addressed, a small portion of such error will pose a significant threat to sub-threshold scaling. For this issue, earlier work has proposed to attach another LRU to deal with a similar problem \cite{brown2020critical} while we deal with this problem from a decoding perspective. The located error has a distance $d_e = d$ \cite{msbj-fxw7}, therefore a \textit{critical fault} has distance $d_e = \frac{d+1}{2}$ if the two-qubit error chain is completely located. That is enough to guarantee the effectiveness of error correction since its distance is the same as the Pauli error. Therefore, all we need to do is to locate the \textit{critical fault}.\\

To decode erasure errors, one possible way is to adapt the weight of edges in the decoding graph representing error mechanisms, according to the information of the erased qubit, based on the MPWM algorithm \cite{Gidney_2021}. The erasure error can be regarded as a completely mixed state, namely independent X and Z errors with 50\% probability \cite{PhysRevLett.102.200501}. The weight $w = \ln\frac{1-p}{p}$ is set to 0 so that the error is located when we try to find a minimum weight perfect matching \cite{wu2023fusionblossomfastmwpm}. Our work is based on the general located error where we only use final leakage detection information to locate the error \cite{d1v7-nctj,msbj-fxw7,Perrin_2025}. When the data qubits are measured in each round, three outcome measurement is applied to distinguish whether each qubit is leaked ($\ket{L}$) or is projected to the eigenstate in the qubit subspace ($\ket{0}$ or $\ket{1}$) \cite{msbj-fxw7,Perrin_2025,suchara2015leakage,PRXQuantum.5.040343,cong2022hardware}. Once a leakage error is detected, we infer the average error probability of each error mechanism. For example, if a data qubit in even line is measured to be leaked, corresponding $D_{1e}(0)$ has $p = \frac{p_e/2}{1-(1-p_e/2)^{10}} \approx \frac{1}{10}$ to happen (See Appendix \ref{appendix2} for details). If one error mechanism is related to multiple leakage detection results, we calculate the total average probability by assuming their occurrences are independent. We use this method to consider all possible error mechanisms and use \textit{pymatching} to decode the re-weighted decoding graph \cite{Higgott2025sparseblossom}.\\

 \begin{figure}[!htbp]
    \centering
    \includegraphics[width=0.48\textwidth]{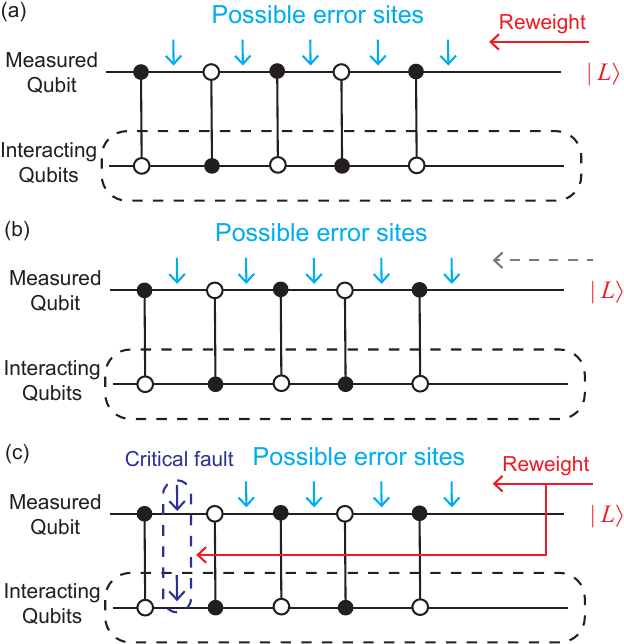}
    \captionsetup{justification=raggedright}
    \caption{Comparison between three decoders. The circuit is an abstract one and does not represent a real circuit. A series of interacting qubits is plotted together, and one measured qubit is drawn out explicitly. The re-weight process needs consideration of all measured qubits. \textbf{(a)} The located decoder accounts for leakage in all error sites on this measured qubit by reweighting their corresponding error mechanism. \textbf{(b)} The trivial decoder does not have a reweighting process. \textbf{(c)} The critical decoder implements the reweighting process in the same way as the located decoder, except for the error site of the critical fault. For the critical fault, we account for the two error sites (one is on the measured qubit shown explicitly, and another is on the qubit that interacts with the measured qubit through the CNOT gate, where the critical fault may happen). In this way, we guarantee that the two-qubit error chain induced by the critical fault is fully located, even when only one type of leakage is detected.}
    \label{fig3}
\end{figure}

\section{Numerical Results}\label{sec3}
In this section, we provide numerical results to demonstrate the performance of our protocol. We consider two conditions and compare three different decoders, including the \textit{located decoder}, the \textit{trivial decoder}, and the \textit{critical decoder}, see Fig.\ref{fig3} for the comparison of the three decoding strategies. The \textit{located decoder} means that we adapt the weight according to the detection of leakage, including atom loss and radiative decay \cite{cong2022hardware}. The \textit{trivial decoder} means that we do not adapt the weight of the decoding graph according to the result of leakage detection, and it is just for comparison.  When we can only distinguish one type of leakage from the qubit subspace, radiative decay or atom-loss, we slightly modify the \textit{Located decoder} to deal with the \textit{critical fault}. The \textit{critical decoder} is motivated by the fact that \textit{correlated leakage} is decay \& atom loss \cite{wu2022erasure}. Detecting only one type of Rydberg decay is enough to address this critical fault, given a proper decoder design.\\

 Our simulation includes Rydberg decay and the two-qubit depolarization error arising from two-qubit gates \cite{wu2022erasure}. Here, the terminology Rydberg decay refers to atom loss error from conversion of BBR and decay error out of the qubit subspace. The decay error back to the qubit subspace, which takes up a small branch of Rydberg decay ($\sim 5\%$), is accounted for by the two-qubit depolarization error \cite{wu2022erasure}. For each CNOT gate, we assume it has $p_e$ probability to have the Rydberg decay error and $p_d$ probability to have the two-qubit depolarization error. The ratio of erasure (Rydberg decay) is defined as $R_e = \frac{p_e}{p_e+p_d} = \frac{p_e}{p}$. For the Rydberg decay error in each CNOT gate, we sample an error operator from $\{\frac{1-\eta}{2} L\otimes P,\frac{1-\eta}{2} P\otimes L,\eta L\otimes L\}$, with a total error rate $p_e$. $P$ represents 50\% Pauli X or Z error according to the end of the qubit (control qubit of CNOT: Z, target qubit of CNOT: X), $L$ represents $\{K_{0L},K_{1L}\}$ with equal probability, and $\eta$ represents the ratio of correlated leakage. $\eta = 0$ represents a condition that the critical fault does not exist and $\eta = 0.0755$ represents a realistic condition \footnote{Estimated for result in Supplementary Information S4 in \cite{wu2022erasure}, the ratio $\eta = \frac{leakage\&leakage}{leakage\&leakage + single-leakage} \approx \frac{P_{BR}}{P_{BR}+P_{QB}+P_{QR}} \approx \frac{\Gamma_R R_{11}}{3(\Gamma_R+\Gamma_B) + \Gamma_Q(R_{11}+R_{11}^{'})/2} = 0.0755$}. The ratio of the correlated leakage varies depending on the atom species and gate sequence, but it should remain a small fraction. For the two-qubit depolarization error, we draw an error operator from $\{I,X,Y,Z\}^{\otimes 2}\setminus{\{I\otimes I\}}$, each with a probability $\frac{p_d}{15}$. The procedure for handling two-qubit depolarization errors is discussed in the Appendix \ref{appendix3}.\\

\begin{figure}[!htbp]
    \centering
    \includegraphics[width=0.48\textwidth]{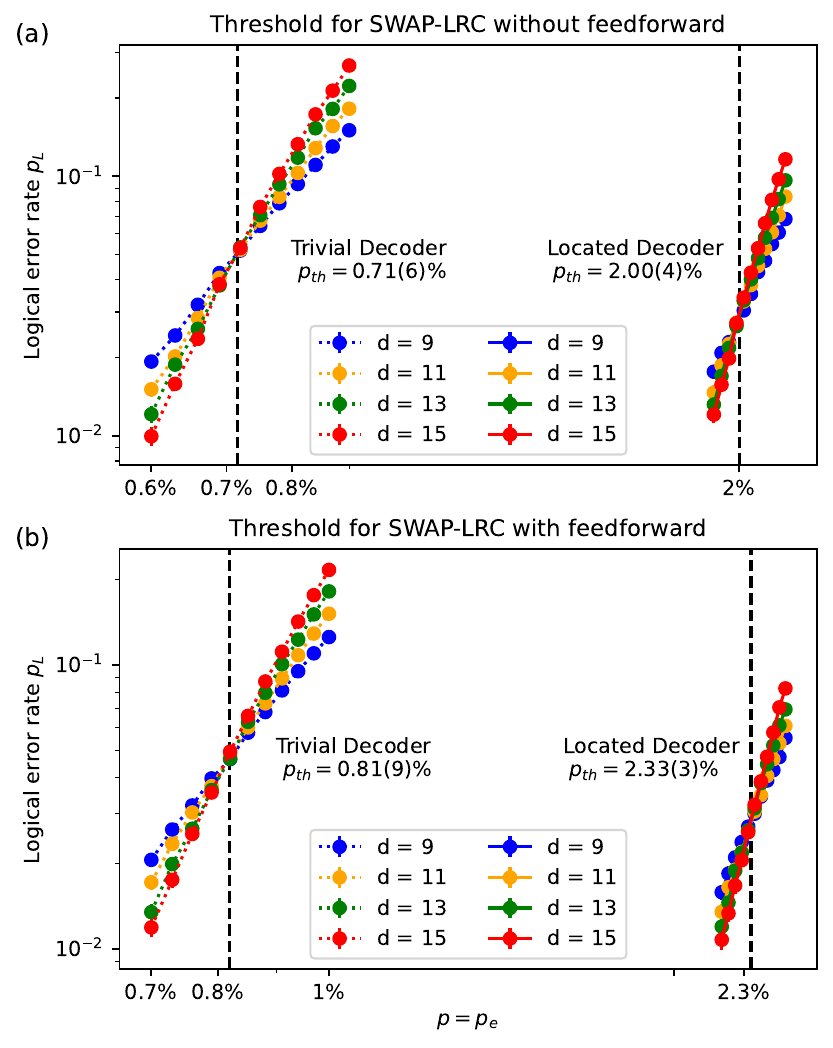}
    \captionsetup{justification=raggedright}
    \caption{Comparison in terms of thresholds between the located decoder and the trivial decoder ($\eta=0.0755$). Dashed lines represent results from the trivial decoder, while solid lines represent results from the located decoder. We sampled $10^5$ shots for each logical error rate. \textbf{(a)} For the circuit without a feed-forward gate, the located decoder has a greatly enhanced threshold ($p_{th} = 2.00(4)\%$) over the trivial decoder ($p_{th} = 0.71(6)\%$). \textbf{(b)} For the circuit with feed-forward gate, the located decoder has threshold $p_{th} = 2.33(3)\%$ and trivial decoder $p_{th} = 0.81(9)\%$.}
    \label{fig4}
\end{figure}

\begin{figure}[!htbp]
    \centering
    \includegraphics[width=0.48\textwidth]{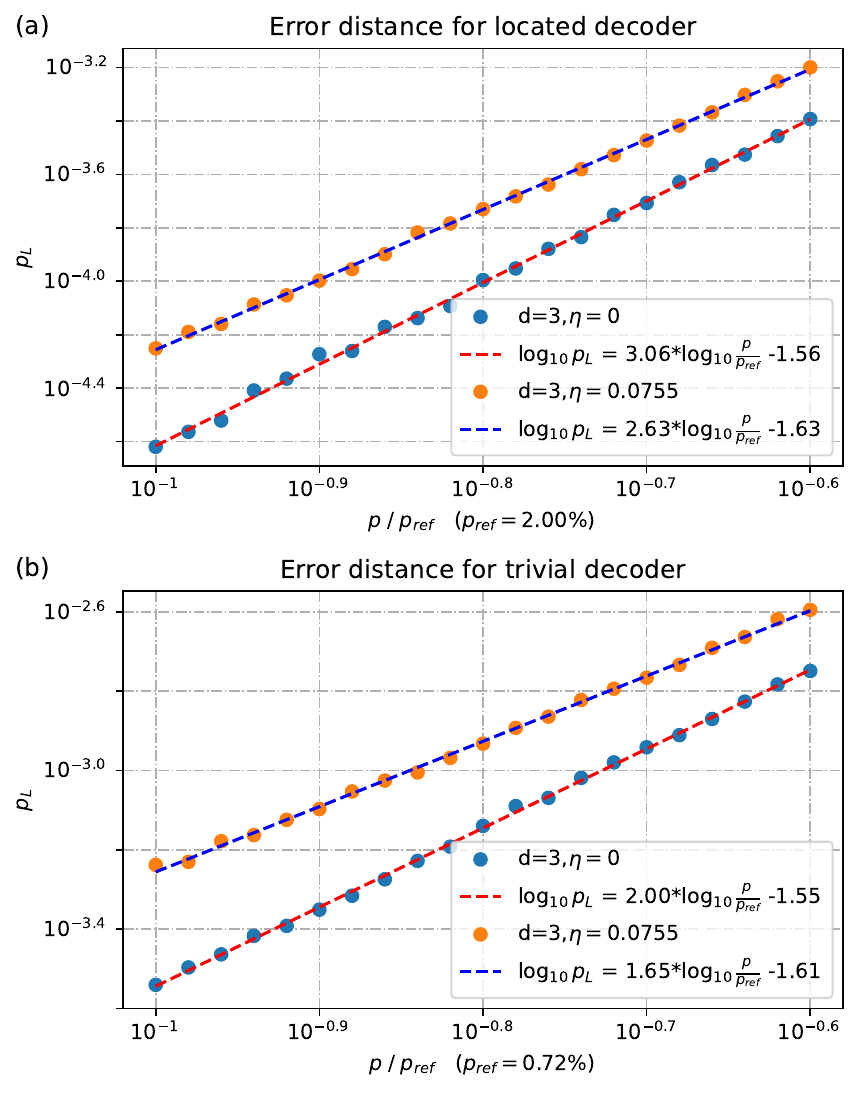}
    \captionsetup{justification = raggedright,singlelinecheck = false}
    \caption{Sub-threshold performance and (effective) error distance. We use the threshold of $\eta=0.0755$ for a reference, and the physical error rate is sampled from $10^{-1}*p_{ref}$ to $10^{-0.6}*p_{ref}$ in logscale. The error distance is derived from the slope of linear regression for $\log_{10} p_{L}$ and $\log_{10} p/p_{ref}$ \textbf{(a)} For the located decoder, the (effective) error distance gives $d_e = 3.06 \approx d$ when $\eta = 0$ and it degrades to $d_e = 2.63$ with a small branch of the \textit{correlated leakage} error ($\eta = 0.0755$). Despite being slightly degraded, the effective error distance remains higher than that of the Pauli error $d_e = \frac{d+1}{2} = 2$. \textbf{(b)} For trivial decoder, the error distance is $d_e = 2.00 \approx \frac{d+1}{2}$ when $\eta = 0$. If the \textit{correlated leakage} error is in the presence, it degrades to $d_e = 1.65 < \frac{d+1}{2}$.}
    \label{fig5}
\end{figure}

 We discuss the performance of the located decoder and the trivial decoder in Sec.\ref{ssec3-1} when both types of Rydberg decay can be detected. We first consider the performance when $R_e = 1$ to show the thresholds for pure Rydberg decay errors and clarify our discussion on the sub-threshold scaling. Then we involve the performance when $R_e \neq 1$ to show the performance advantages compared to traditional error correction protocols. We then discuss the performance of three decoders in Sec.\ref{ssec3-2}, when only one type of Rydberg decay is detected. We show that \textit{critical decoder} effectively eliminates the detrimental effects of Rydberg decay with minimal hardware requirements. In most of our results, we use the circuit in Fig.\ref{fig1}(b), but we also benchmark the threshold of the circuit in Fig.\ref{fig1}(c) in Fig.\ref{fig4} \cite{Baranes_2026}. The logical error rate only accounts for longitudinal operator ($X_2$ in Fig.\ref{figS2}) when considering error distance (Fig.\ref{fig5},Fig.\ref{fig7}). Otherwise, it accounts for both operators (A logical error is recorded as long as one of the logical operator measurements is violated). The logical error rates with deviation are estimated through \textit{proportion\_confint} function \cite{seabold-proc-scipy-2010} in \textit{statsmodels.stats.proportion}, by calculating $99\%$ confidence interval. The code is available in \cite{code}.\\
 
\subsection{Condition I: Both Types Detected (Atom loss and Decay error)}\label{ssec3-1}
Here, we consider the condition that both types of Rydberg decay can be detected. This condition requires detecting the two types of Rydberg decay. For alkaline earth atoms such as $\ce{^{171}_{}Yb}$, this requires that we first detect the decay error to the ground state and then implement three outcome measurements to distinguish states in the qubit subspace and atom loss \cite{wu2022erasure,ma2023high,msbj-fxw7}. For alkali metal atom such as $\ce{^{87}_{}Rb}$, we may image the atoms with the aid of a Stern-Gerlach magnetic field gradient in time-of-flight in principle \cite{PhysRevApplied.19.034089} or we may add leakage detection unit (LRU) to detect the Rydberg decay error \cite{PRXQuantum.5.040343,cong2022hardware}.\\

\begin{figure*}[!htb]
    \centering
    \includegraphics[width=\textwidth]{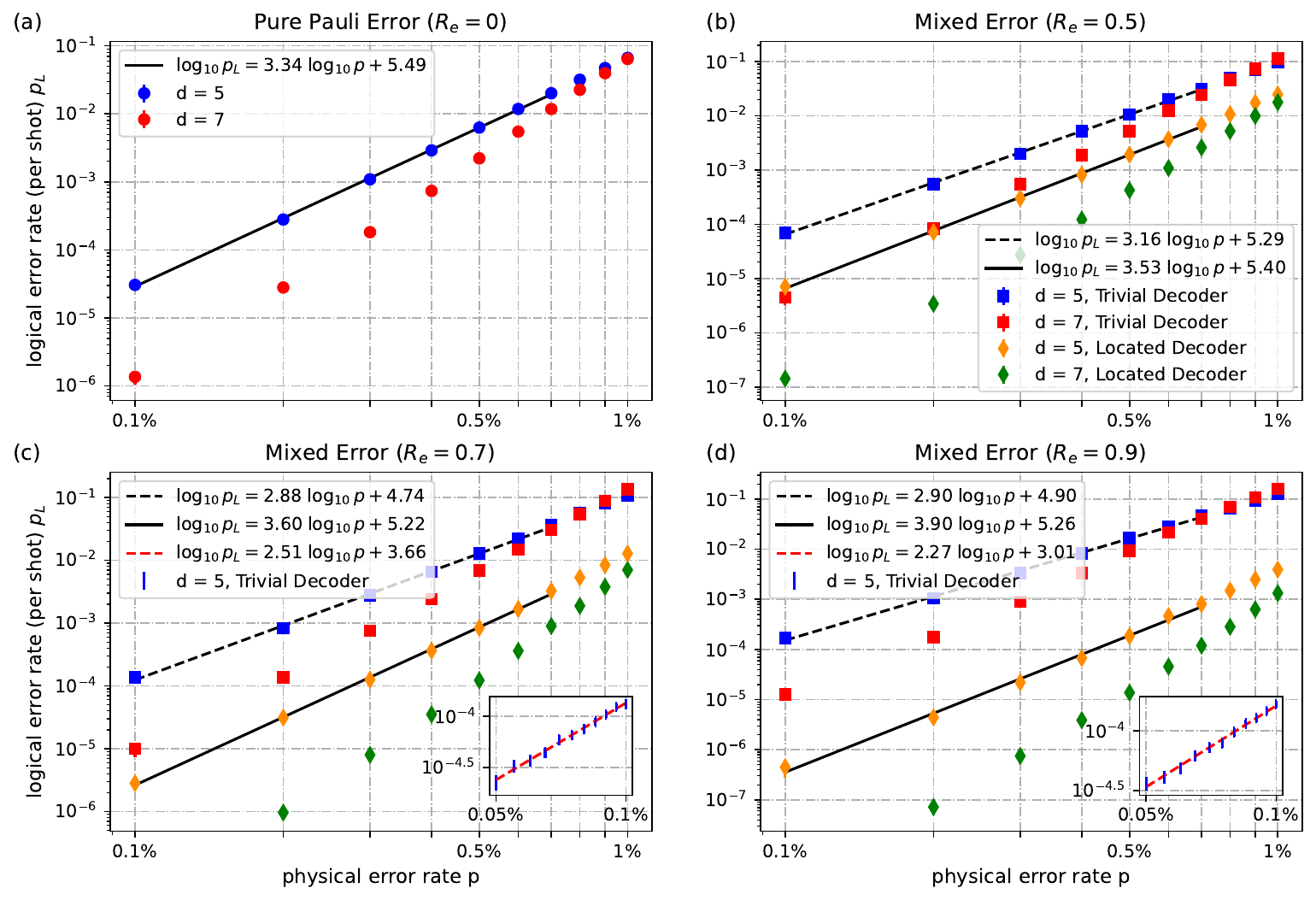}
    \captionsetup{justification = raggedright,singlelinecheck = false}
    \caption{Comparison in the performance of the mixed error. The effective error distance is derived from the slope of linear regression for the logical error rate and the physical error rate below threshold in logscale, for $d=5$. We use a numerical technique used in \cite{985g-58gd} that we only sample a part of leakage samples (2000 leakage samples for $R_e = 0.5,0.7$ and 20000 leakage samples for $R_e = 0.9$). These results indicate that the located decoder has an advantage in sub-threshold scaling while the trivial decoder has a disadvantage. The advantage or disadvantage enlarges when $R_e$ increases. \textbf{(a)} For pure Pauli error, $d_e = 3.34$. This is slightly larger than $d_e = \frac{d+1}{2}$ because the considered physical error rate is not small enough. So this error distance only serves as a reference for comparison \textbf{(b)} $R_e = 0.5$, $d_e = 3.16 < 3.34$ for trivial decoder and $d_e = 3.53 > 3.34$ for located decoder; \textbf{(c)} $R_e = 0.7$, $d_e = 2.88 < \frac{d+1}{2}$ for trivial decoder and $d_e = 3.6 > 3.34$ for located decoder. The inset figure is the result for a smaller physical error rate; \textbf{(d)} $R_e = 0.9$, $d_e = 2.9 < \frac{d+1}{2}$ for the trivial decoder and $d_e = 3.9 > \frac{d+1}{2}$ for the located decoder. We have included results for a smaller physical error rate because the considered physical error range does not reveal the change in sub-threshold scaling for the trivial decoder. }
    \label{fig6}
\end{figure*}

The result of the threshold for the pure Rydberg decay ($R_e = 1$) is shown in Fig.\ref{fig4}. We define one shot as performing $d$ rounds of syndrome measurements on a toric code of distance $d$ using SWAP-LRC. Both types of circuits are considered here, and the circuit with a feed-forward gate achieves higher thresholds because this circuit has fewer CNOT gates. We derive the threshold as a critical point of phase transition from universal scaling ansatz with data sets close to the threshold regime for $d = 9,11,13,15$ \cite{Wang_2003}. Fig.\ref{fig4} shows that the located decoder has a greatly enhanced threshold over trivial decoder ($2.0\%$ vs $0.72\%$ for circuit without feed-forward and $2.33\%$ vs $0.82\%$ for circuit with feed-forward) and it also has clear advantage over the traditional Pauli error ($0.937\%$ for XZZX surface code without SWAP-LRC \cite{wu2022erasure}).\\

The result of the sub-threshold performance and the error distance for pure Rydberg decay is shown in Fig.\ref{fig5}. A condition without the correlated leakage ($\eta=0$) is also illustrated for comparison. The result shows that the effective error distance is highly sensitive toward the lower error distance $d_e = \frac{d+1}{2}$. In Fig.\ref{fig5}(a), when a located decoder is used, the effective error distance $d_e = 2.63$ still has an apparent advantage over the traditional Pauli error $d_e = \frac{d+1}{2} = 2$ since the correlated leakage only takes up a small fraction. This suggests better suppression of logical error with a located decoder. But for the trivial decoder (Fig.\ref{fig5}(b)), the distance is $d_e = \frac{d+1}{2}$ error without the \textit{critical fault} ($\eta = 0$). Once it exists, even a small fraction of the \textit{correlated leakage} error degrades the distance and threatens the sub-threshold scaling. \\

We next consider a mixture of the Pauli error and the Rydberg decay error. In neutral atoms, Pauli errors come from not only Rydberg decay back to the qubit-subspace but experimental imperfections as well, such as dephasing from imperfect laser pulses and atom heating. Here we select $R_e = 0.5,0.7,0.9$ and compare their performance with pure Pauli (two-qubit depolarization) error. We select an error range from $0.1\%$ to $1\%$, and each point is sampled for enough shots to suppress the deviation.\\

The results in Fig.\ref{fig6} reveal two things. For the threshold, the Rydberg decay error has slightly lower thresholds with \textit{trivial decoder} compared to the threshold of the two-qubit depolarization error, but has higher thresholds with \textit{located decoder}. For the sub-threshold scaling, the Rydberg decay error has slightly lower effective error distances with \textit{trivial decoder} compared to that of the Pauli error because of \textit{critical fault}, but has higher distance with the \textit{located decoder} since a large fraction of errors induced by the Rydberg decay error are located with $d_e = d$. The difference (especially in the sub-threshold scaling) enlarges when $R_e$ increases. \\

We notice that the considered error rate is not small enough to show the changes in error distances for the trivial decoder when $R_e$ changes from $0.7$ to $0.9$, so we have extended the considered physical error range to $0.05\%$. The results for the extended physical error range also emphasize that even if the ratio of \textit{critical fault} is rather small, it degrades the performance of the sub-threshold scaling greatly when the physical error rate is small enough. This reveals the importance of addressing the harmful critical error with an error distance $d_e = \lfloor\frac{d+3}{4}\rfloor$ to maintain the performance of error correction. At the same time, the gap in logical error rates between three decoders also emphasizes the necessity of the located decoder, especially when $R_e$ is large. When $R_e = 0.9$ and $p<0.5\%$, the logical error rates of the located decoder outperform those of the trivial decoder by two orders of magnitude.

\subsection{Condition II: One Type Detected (Atom loss or Decay error)}\label{ssec3-2}

Here, we consider the condition that only one branch of Rydberg decay is detected and introduce the \textit{critical decoder} that is suitable for a more experimental-friendly implementation. This condition is of special interest because it is relatively inconvenient to detect the decay error into hyperfine levels for alkali metal atoms now, but atom loss from anti-trapping of atoms in Rydberg state can be detected with the existing technique \cite{PRXQuantum.5.040343,evered2025probingtopologicalmatterfermion}. Although the measurement in \cite{PhysRevApplied.19.034089}, which images the atom ensemble with the help of a Stern-Gerlach magnetic field gradient in time-of-flight, is effective in detecting both types of Rydberg decay, it has not become a standard technique for the individual atom. For alkaline-earth atoms, it is also easier to detect one type of leakage than to detect both types \cite{ma2023high}. The purpose of the critical decoder does not lie in improved performance over the traditional Pauli error, since this advantage needs a large branch of error to be detected, which is difficult to realize when we only detect one type of leakage \cite{wu2022erasure}. Instead, we aim to find an efficient protocol to mitigate the damaging effect on the sub-threshold scaling from the decoding perspective. Since the \textit{critical fault} that degrades the distance is the correlated leakage during the first gate and it must be a decay error alone with atom loss, we assume that the critical decoder is enough to locate the most harmful error, by converting its distance from $d_e = \lfloor \frac{d+3}{4} \rfloor$ to $d_e = \lfloor \frac{d+1}{2} \rfloor$. So it serves as a hardware-efficient method to address Rydberg decay.\\

\begin{figure}[!htbp]
    \centering
    \includegraphics[width=0.48\textwidth]{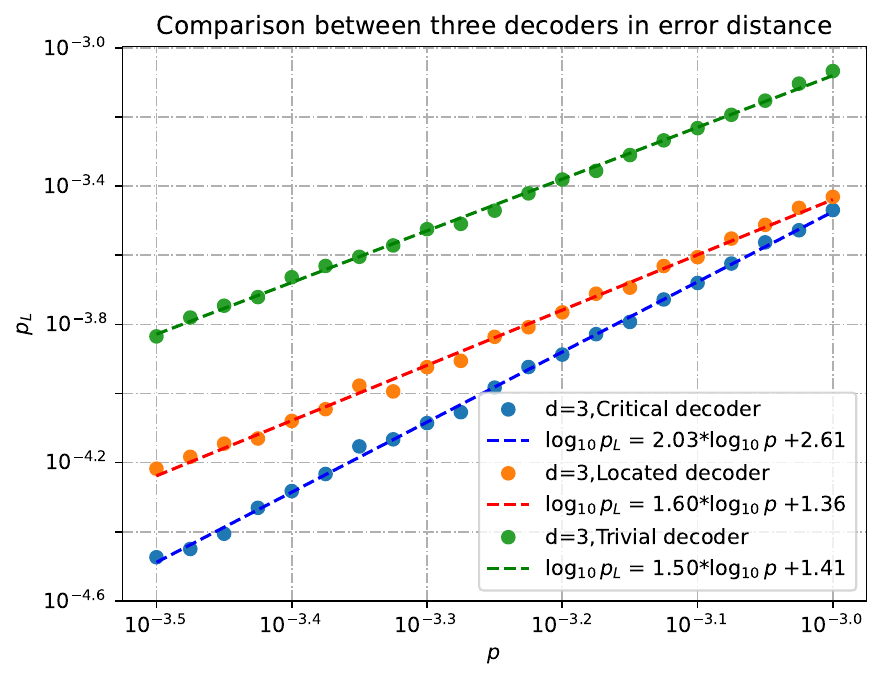}
    \captionsetup{justification = raggedright,singlelinecheck = false}
    \caption{Comparison between three decoders when only one type of Rydberg decay can be detected. We set $R_e = 0.9$ and the ratio of the detected leakage to be $50\%$. Only the critical decoder preserves the distance.}
    \label{fig7}
\end{figure}

In our critical decoder (see Fig. \ref{fig3}(c)), we assume the undetected leakage is measured to be 0/1 with equal probability. To account for the critical fault, we assume that the correlated leakage is bound to occur after the first CNOT gate in our decoder, instead of occurring with probability proportion to $\eta$. We then reweight the edges representing data qubit errors according to this specific assignment. This strategy over-estimates its probability, but it guarantees that we can locate the critical fault in decoding. We compare the three decoders when $R_e = 0.9$ and the ratio of detected leakage is $50\%$. \\

The result is shown in Fig.\ref{fig7}. It reveals that only the \textit{critical decoder} preserves the distance. The difference in effective error distance lies in what kinds of errors a \textit{critical fault} introduces. For the \textit{trivial decoder}, a critical fault introduces two non-located errors along one logical operator, so the distance is degraded. For the \textit{located decoder}, it introduces one located error and one non-located error. This condition is less harmful than the condition above, but it still degrades the distance \footnote{For $d = 3,5$, one or two critical faults are enough to introduce a logical error with both trivial decoder and located decoder. For $d = 7$, three critical faults are needed to introduce a logical error with a located decoder, while two critical faults are enough to introduce a logical error with a trivial decoder. For all conditions, the error distance is lower than $d_e = \frac{d+1}{2}$}. One \textit{critical fault} is enough to introduce a logical error in a $d = 3$ code. However, with the \textit{critical decoder}, a critical fault introduces two located errors. Since a code with distance $d$ corrects $d-1$ located errors, at least $d_e = \frac{d+1}{2}$ gate errors are required to introduce a logical error. The error distance for the critical fault is the same as the traditional Pauli error, so it does not introduce additional harm to the sub-threshold scaling.

\section{Discussions and Conclusion}\label{sec4}
In conclusion, our work provides a hardware-efficient protocol to address Rydberg decay. Neither additional hardware operations nor additional ancilla qubits or CNOT gates are required with our strategy, unlike erasure conversion in alkaline-earth metal atoms \cite{wu2022erasure} and previous work on \textit{Partial-LRU} \cite{suchara2015leakage,Perrin_2025}. The only requirement or overhead is the final three-outcome measurement, which distinguishes the leaked state from the qubit subspace to locate the error \cite{cong2022hardware,suchara2015leakage,PRXQuantum.5.040343}. The three-outcome measurement has been demonstrated recently, which indicates that our protocol is practical in neutral atoms \cite{PRXQuantum.5.040343}. For pure Rydberg decay, we show a high threshold $2.33\%$. For a more realistic condition where Pauli error exists. We demonstrate that Rydberg decay is not only harmless but also has enhanced performance compared to traditional Pauli error, owing to the property of located error. Furthermore, we examine the scenario where only one type of leakage is detectable - a condition of practical interest \cite{PRXQuantum.5.040343,evered2025probingtopologicalmatterfermion,ma2023high}. We develop the \textit{critical decoder} motivated by the property of the \textit{critical fault} \cite{brown2020critical} and the located error \cite{msbj-fxw7}, and we show it is enough to eliminate the damaging effect of the critical fault in sub-threshold scaling and preserves the distance $d_e = \frac{d+1}{2}$ \cite{jandura2024surfacecodestabilizermeasurements}. Our strategy is more efficient compared to the previous protocol that used an additional LRU to remove the critical fault \cite{brown2020critical}.

For future work, we first expect that our method to address Rydberg decay can be extended to other types of leakage in different hardware systems, such as atom loss in neutral atoms \cite{Baranes_2026,Perrin_2025}, leakage in an ion-trap system \cite{PhysRevA.100.032325} and leakage in dual-rail qubits \cite{PhysRevX.13.041022,d1v7-nctj}. We also notice that in SWAP-LRC, Rydberg decay inevitably introduces time-correlated error. Such an error is common in the process of correcting errors involving leakage \cite{brown2020critical,PhysRevA.100.032325} and it may need constant times of syndrome extraction rounds to suppress in lattice-surgery \cite{timelike,Gidney_2022}. But its effect on transversal gate and correlated decoding strategy is not clear yet \cite{PhysRevLett.133.240602}. Therefore, an important direction is to study the performance of such a decoding strategy in logical algorithms \cite{Baranes_2026}.\\
\textit{Note.—A parallel study \cite{Baranes_2026} also examines atom loss in fault-tolerant algorithms using a comparable decoding approach. However, our work is distinguished by its distinct scope and technical details. We highlight that Ref. \cite{Baranes_2026} proposes an optimized SWAP-LRC circuit—incorporating feed-forward gates and virtual corrections—and evaluates its error threshold in Fig. \ref{fig4}.} \\

\textit{Acknowledgments}---We acknowledge helpful discussions with Zihan Chen, Yuchen Zhang, Zhaoqiu Zengxu, Siyuan Chen, Pai Peng, Gefen Baranes and Pengyu Liu. This work was supported by the HFNL Self-Deployed Project (Nos. ZB2024010101, ZB2024010102, ZB2024010201, and ZB2024010501), the Quantum Science and Technology-National Science and Technology Major Project (No. 2021ZD0301405), the National Key R\&D Program of China (No. 2022ZD0160101), the Shanghai Municipal Science and Technology Commission (No. 24DP2600300), the National Natural Science Foundation of China (No. 12322415), and the New Cornerstone Science Foundation.

\appendix
\onecolumngrid
\section{Derivation of the Error Propagation Property}\label{appendix1}

In this appendix, we derive the error propagation property of Rydberg decay \cite{msbj-fxw7}. First, we briefly review our approach to address the Rydberg decay error. With two low-lying levels to encode a qubit and only $\ket{1}$ is strongly coupled with the Rydberg state, the channel of the Rydberg decay error is described in the operator-sum form with two Kraus operators $\xi(\rho) = \sum_{i=0,1} K_i \rho K_i^{\dagger}$ \cite{sahay2023high,cong2022hardware}. The Kraus operators are given by:
\begin{equation}
\begin{cases}
K_0 = \ket{0}\bra{0} + \sqrt{1-p_e}\ket{1}\bra{1} + \ket{L}\bra{L}\\
K_1 = \sqrt{p_e}\ket{L}\bra{1}
\end{cases}
\label{eqn1}
\end{equation}
We refer to $K_0$ as the no-jump operator and $K_1$ as the jump operator. We don't consider the jump operators between qubit subspaces that are modeled as Pauli errors with Pauli twirling approximation (PTA), because Pauli errors are compatible with traditional error correction schemes. For radiative decay (RD), the leaked state $\ket{L}$ is energetically separated from the qubit subspace (For alkali metal atoms, an additional magnetic field is required). For blackbody radiation (BBR) or other mechanisms that lead to residual Rydberg population, the leaked state $\ket{L}$ represents the lost atom because the atoms in the Rydberg state are ejected by the anti-trapping potential. Under both conditions, the leaked state $\ket{L}$ is not involved in subsequent single- or two-qubit gates, so our scheme dealt with RD and BBR uniformly. \\

As pointed out previously \cite{msbj-fxw7}, when we apply Pauli twirling and randomized compiling in the presence of the leaked state $\ket{L}$, non-diagonal terms in the process matrix are removed for the no-jump evolution as usual, while the jump evolution governed by $K_1 \rho K_1^{\dagger}$, representing biased erasure channel, is converted to the erasure channel governed by two Kraus operators as below ($\overline{\xi_1(\rho)}^{P'} = \sum_{j = 0,1} K_{jL} \rho K_{jL}^{\dagger}$).

\begin{equation}
    \begin{cases}
        K_{0L} = \sqrt{p_e/2}\ket{L}\bra{0}\\
        K_{1L} = \sqrt{p_e/2}\ket{L}\bra{1}
    \end{cases}
\label{eqnA2}
\end{equation}

\begin{figure*}[h]
\renewcommand{\thefigure}{S1}
    \centering
    \includegraphics[width=\textwidth]{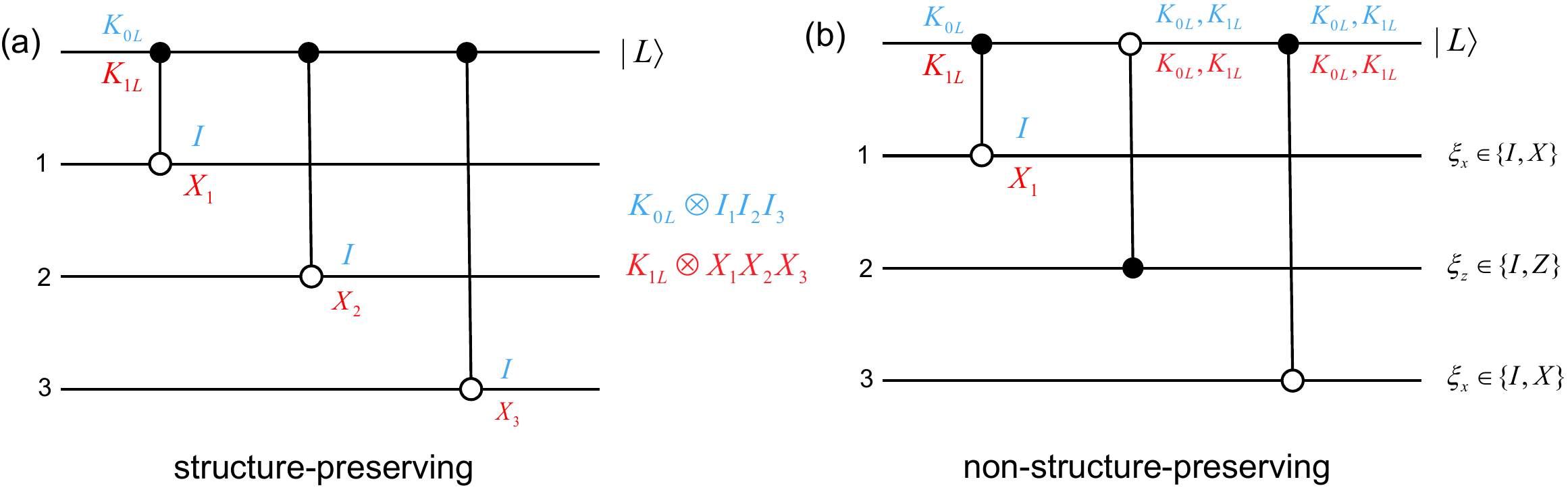}
    \captionsetup{justification = raggedright,singlelinecheck = false}
    \caption{Two different forms of circuits and their error propagation. We consider the jump evolution for the uppermost qubit. (a) The form of the jump operator is preserved when the leaked qubit is always the control of CNOT gates. Therefore, if the jump operator is $K_{1L}$, it propagates three X errors to corresponding qubits, namely a correlated X error. The propagation acts similarly to that of Pauli error, namely, an error propagates to a (correlated) Pauli error through the Clifford gates. (b) The form of the jump operator is not preserved when the leaked qubit is not always the control or target of CNOT gates. We prove that the leaked qubit propagates to 50\% X or Z error to corresponding qubits.}
    \label{figS1}
\end{figure*}

When the leaked qubit is in the control of CNOT gates, $K_{0L}$ commutes with the CNOT gate, and $K_{1L}$ propagates to an X error to the interacting qubit. Therefore, when the leaked qubit is always the control, the form of the jump operator is preserved. $K_{1L}$ propagates to a correlated $X$ error and or $K_{0L}$ does not propagate an error, as shown in Fig.\ref{figS1}(a). The above argument is the same if the leaked qubit is always the target of CNOT gates, by altering the basis of Kraus operator (from $\{\ket{0},\ket{1}\}$ in eqn \ref{eqnA2} to $\{\ket{+},\ket{-}\}$). However, if the condition above is not satisfied, the form of the Kraus operator is not preserved. For example, a $K_{1L}$ in the target of a CNOT gate generates both $K_{1L}$ and $K_{0L}$ as shown below ($CNOT = (I'\otimes I' - \ket{11}\bra{11} - \ket{10}\bra{10} + \ket{10}\bra{11} + \ket{11}\bra{10})$ and $I',X',Y',Z'= I,X,Y,Z\oplus \ket{L}\bra{L}$.)

\begin{equation}
\begin{aligned}
CNOT \; I'\otimes K_{1L} \; CNOT &= \sqrt{p_e/2} \; CNOT(\ket{0L}\bra{01}+\ket{1L}\bra{11}+\ket{LL}\bra{L1})CNOT\\
&= \sqrt{p_e/2}(\ket{0L}\bra{01}+\ket{1L}\bra{10}+\ket{LL}\bra{L1})\\
&= \sqrt{p_e/2}(\ket{0}\bra{0}+\ket{L}\bra{L})\otimes\ket{L}\bra{1} +\ket{1}\bra{1}\otimes\ket{L}\bra{0}\\
&= \frac{1}{2}(I'+Z')\otimes K_{1L}+\frac{1}{2}(I'-Z')\otimes K_{0L}\\
&= \frac{1}{\sqrt{2}} I'\otimes K_{+L} - \frac{1}{\sqrt{2}} Z'\otimes K_{-L}
\end{aligned}
\end{equation}

For the condition in Fig.\ref{figS1}(b), the forward propagation of operator $K_{0L}$ and $K_{1L}$ are given as below (The right arrow represents forward propagation, namely an operator $K$ before some unitary gate $U$ equals to $UKU^{\dagger}$ after this gate.)

\begin{equation}
\begin{aligned}
K_{1L} &\xrightarrow{1st\;gate} X'_1 \otimes K_{1L}\\
&\xrightarrow{2nd\;gate} \frac{1}{\sqrt{2}}(X'_1 I'_2\otimes K_{+L}-X'_1 Z'_2\otimes K_{-L})\\
&= \frac{1}{2}(X'_1 I'_2\otimes K_{0L} + X'_1 I'_2\otimes K_{1L} - X'_1 Z'_2\otimes K_{0L} + X'_1 Z'_2\otimes K_{1L})\\
&\xrightarrow{3rd\;gate}\frac{1}{2}(X'_1 I'_2 I'_3\otimes K_{0L} + X'_1 I'_2 X'_3\otimes K_{1L} - X'_1 Z'_2 I'_3\otimes K_{0L} + X'_1 Z'_2 X'_3\otimes K_{1L})
\end{aligned}
\end{equation}
\begin{equation}
\begin{aligned}
K_{0L} &\xrightarrow{1st\;gate} I'_1 \otimes K_{0L}\\
&\xrightarrow{2nd\;gate} \frac{1}{\sqrt{2}}(I'_1 I'_2\otimes K_{+L}+I'_1 Z'_2\otimes K_{-L})\\
&= \frac{1}{2}(I'_1 I'_2\otimes K_{0L} + I'_1 I'_2\otimes K_{1L} + I'_1 Z'_2\otimes K_{0L} - I'_1 Z'_2\otimes K_{1L})\\
&\xrightarrow{3rd\;gate}\frac{1}{2}(I'_1 I'_2 I'_3\otimes K_{0L} + I'_1 I'_2 X'_3\otimes K_{1L} + I'_1 Z'_2 I'_3\otimes K_{0L} - I'_1 Z'_2 X'_3\otimes K_{1L})
\end{aligned}
\end{equation}

The resulting noisy channel gives

\begin{equation}
\begin{aligned}
\sum_{j = 0,1} K_{jL} \rho K_{jL}^{\dagger} \xrightarrow{3 \; gates} &\frac{1}{4}( X'_1 I'_2 I'_3 \; \rho_{123} \; X'_1 I'_2 I'_3 \otimes K_{0L} \; \rho_{lea} \; K_{0L}^{\dagger} + X'_1 I'_2 X'_3 \; \rho_{123} \; X'_1 I'_2 X'_3 \otimes K_{1L} \; \rho_{lea} \; K_{1L}^{\dagger}\\
& + X'_1 Z'_2 I'_3 \; \rho_{123} \; X'_1 Z'_2 I'_3 \otimes K_{0L} \; \rho_{lea} \; K_{0L}^{\dagger} + X'_1 Z'_2 X'_3 \; \rho_{123} \; X'_1 Z'_2 X'_3 K_{1L} \; \rho_{lea} \; K_{1L}^{\dagger}\\ 
& + I'_1 I'_2 I'_3 \; \rho_{123} \; I'_1 I'_2 I'_3 \otimes K_{0L} \; \rho_{lea} \; K_{0L}^{\dagger} + I'_1 I'_2 X'_3 \; \rho_{123} \; I'_1 I'_2 X'_3 \otimes K_{1L} \; \rho_{lea} \; K_{1L}^{\dagger}\\ 
& + I'_1 Z'_2 I'_3 \; \rho_{123} \; I'_1 Z'_2 I'_3 \otimes K_{0L} \; \rho_{lea} \; K_{0L}^{\dagger} + I'_1 Z'_2 X'_3 \; \rho_{123} \; I'_1 Z'_2 X'_3 \otimes K_{1L} \; \rho_{lea} \; K_{1L}^{\dagger})\\
= \; \; \; & \frac{p_e}{16}(X'_1 I'_2 I'_3 \; \rho_{123} \; X'_1 I'_2 I'_3 + X'_1 I'_2 X'_3 \; \rho_{123} \; X'_1 I'_2 X'_3 + X'_1 Z'_2 I'_3 \; \rho_{123} \; X'_1 Z'_2 I'_3 + X'_1 Z'_2 X'_3 \; \rho_{123} \; X'_1 Z'_2 X'_3\\
\; \; \; & I'_1 I'_2 I'_3 \; \rho_{123} \; I'_1 I'_2 I'_3 + I'_1 I'_2 X'_3 \; \rho_{123} \; I'_1 I'_2 X'_3 + I'_1 Z'_2 I'_3 \; \rho_{123} \; I'_1 Z'_2 I'_3 + I'_1 Z'_2 X'_3 \; \rho_{123} \; I'_1 Z'_2 X'_3) \otimes \ket{L}\bra{L}
\end{aligned}
\end{equation}

In the first right arrow, we have dropped all non-diagonal terms because of the Pauli twirling in qubits $1,2,3$. In the equation, we have make use of $K_{iL} \rho K_{iL}^{\dagger} = \frac{p_e}{2} \rho_{ii} \ket{L}\bra{L}$ ($i = 0,1$). $\rho_{ii}$ is dependent on the density matrix before the leakage. During randomized compiling, the ensemble is a completely mixed state on average, so $\rho_{00} = \rho_{11} = \frac{1}{2}$. Therefore, the leaked qubit propagates to independent 50\% X or Z error to corresponding qubits, namely a kind of tailored Pauli propagation \cite{d1v7-nctj}.
\section{Error propagation of SWAP-LRC}\label{appendix2}

In this appendix, we consider the error propagation in detail for the toric code with SWAP-LRC. Firstly, a schematic diagram of the toric code (Fig.\ref{figS2}) is shown below. After each round of the syndrome measurement, the ancilla qubit for the syndrome measurement becomes the data qubit for the next round. Data qubits in odd lines exchange with ancilla qubits for Z syndrome measure right below them, and data qubits in even lines exchange with ancilla qubits for X syndrome measure right below them. 

\begin{figure*}[h]
\renewcommand{\thefigure}{S2}
    \centering
    \includegraphics[width=0.3\textwidth]{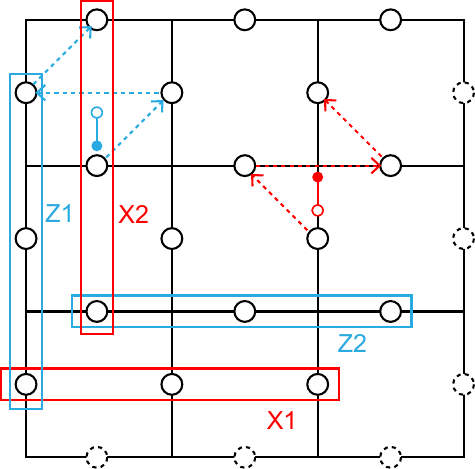}
    \captionsetup{justification = raggedright,singlelinecheck = false}
    \caption{Toric code (for $d=3$): A toric code with distance $d$ has $2d^2$ data qubits (circles on edges), $d^2$ Z stabilizers (plaquette operator) and $d^2$ X stabilizers (site operator). Dashed circles represent a period boundary condition. Two sets of logical operators are labeled with solid rectangles. The gate sequence is labeled with dashed arrows.}
    \label{figS2}
\end{figure*}

\begin{figure*}[h]
\renewcommand{\thefigure}{S3}
    \centering
    \includegraphics[width=0.8\textwidth]{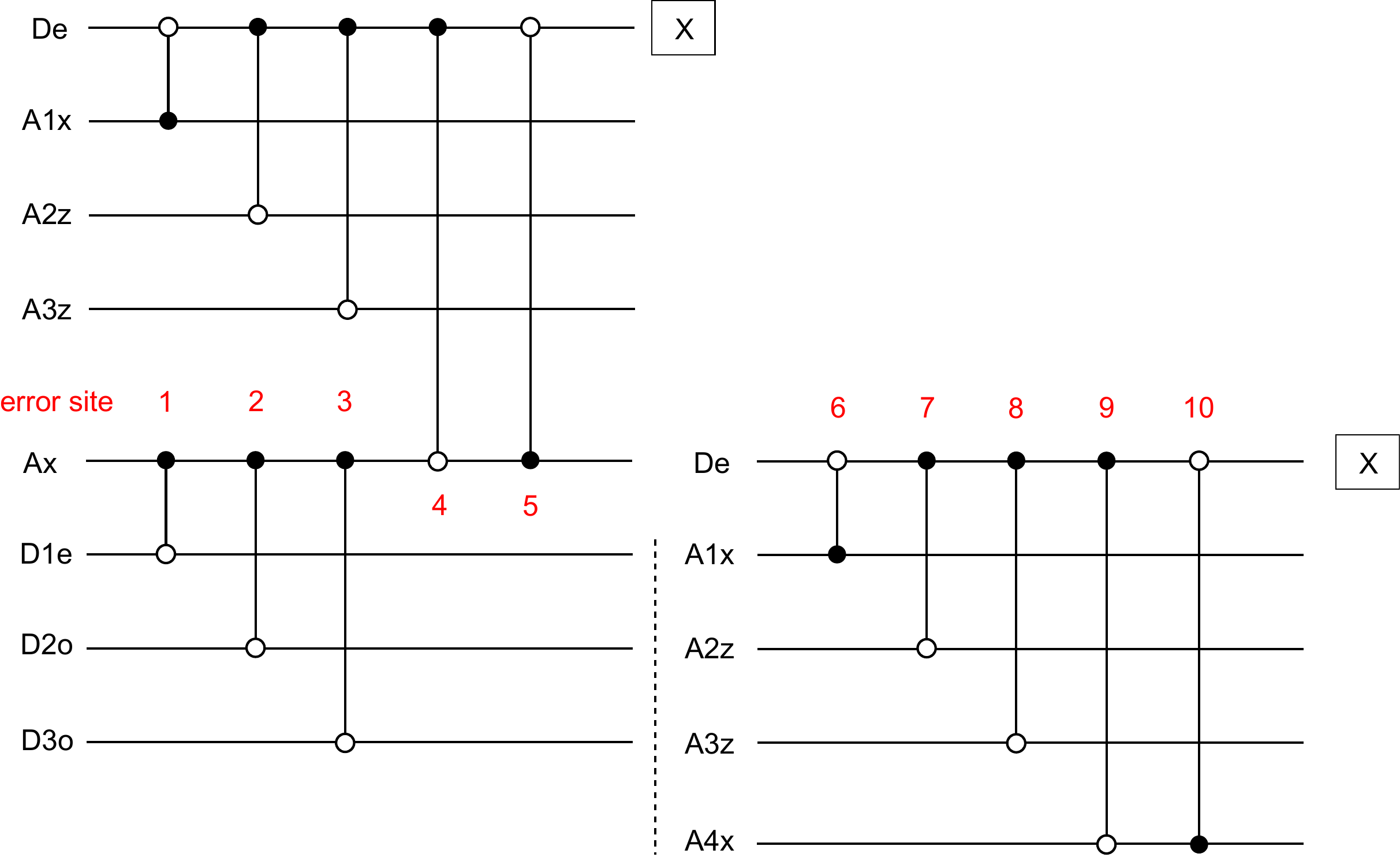}
    \captionsetup{justification = raggedright,singlelinecheck = false}
    \caption{Possible error sites for data qubits in even lines and X ancilla qubits: Each round contains five CNOT gates, and ancilla leakage is detected in the next round of the syndrome measurement, so there are ten nonequivalent error sites. D/A represent the data qubit or the ancilla qubit; x/z represent the type of the ancilla qubit, and e/o represent the type of the data qubit (in even lines or odd lines); the number represents the related qubit in the corresponding time sequence (we assume the fourth and fifth CNOT gate is implemented in the 4th sequence). When considering a circuit with the feed-forward gate, the fifth gate is replaced by a feed-forward gate and is virtually implemented with software correction.}
    \label{figS3}
\end{figure*}

\begin{figure*}[h]
\renewcommand{\thefigure}{S4}
    \centering
    \includegraphics[width=0.8\textwidth]{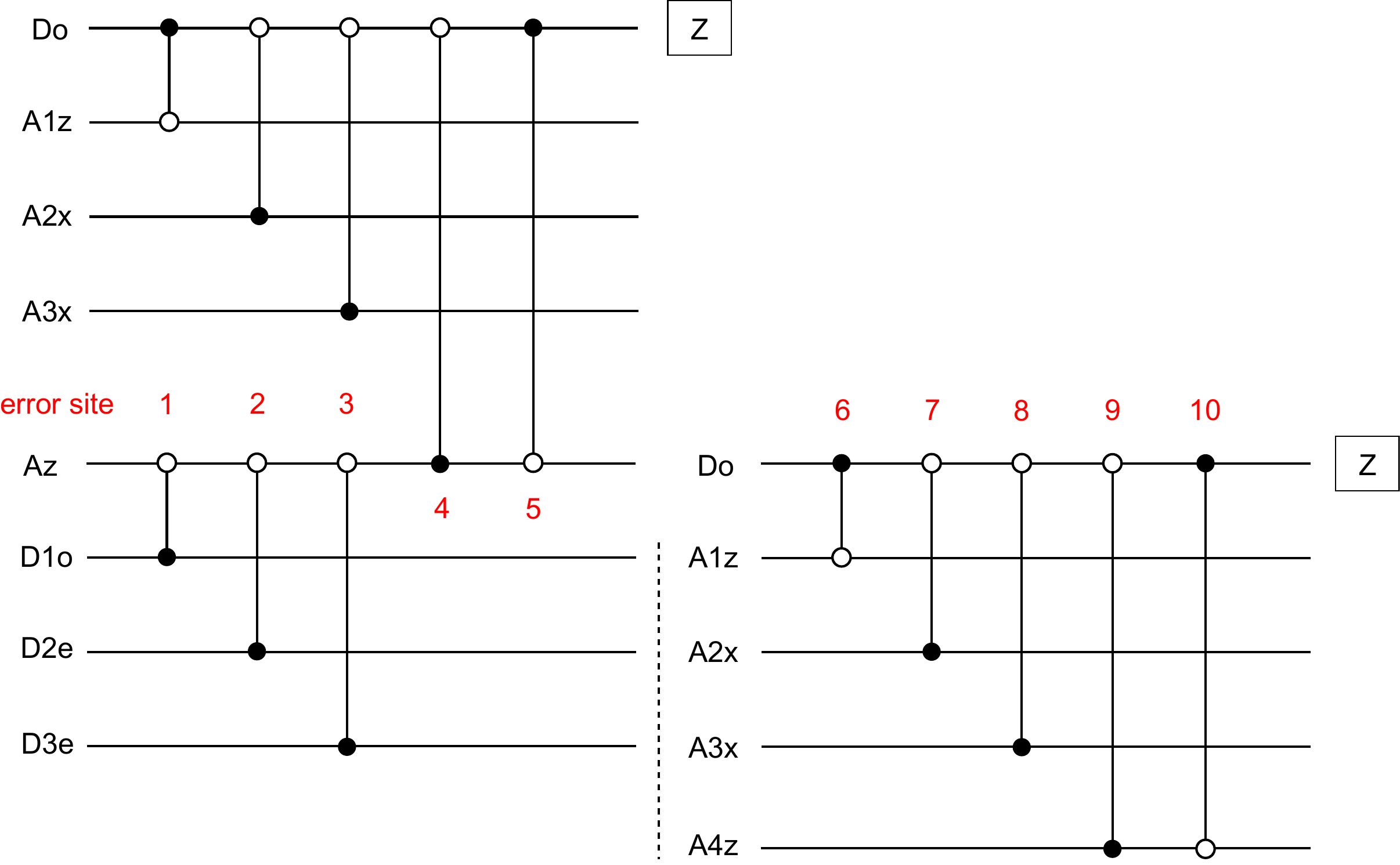}
    \captionsetup{justification = raggedright,singlelinecheck = false}
    \caption{Possible error sites for data qubits in odd lines and Z ancilla qubits}
    \label{figS4}
\end{figure*}

We have plotted all possible error sites, related CNOT gates, and qubits if the measured data qubit is leaked. Data qubits in even lines are shown in Fig. \ref{figS3} and data qubits in odd lines are shown in Fig. \ref{figS4}. In our consideration, we only consider logical X error because logical Z error is derived in the same way. Therefore, we need to consider X errors (or leakage) in data qubits and Z syndrome measurement error (or leakage). We assume the syndrome measurement is repeated for $d$ rounds. In the final round of measurement, data qubits are measured ideally, together with the information on whether the qubit is leaked.

\begin{minipage}[c]{\textwidth}
\centering
\renewcommand\arraystretch{1.5}
\renewcommand{\thetable}{A1}
\begin{tabular}{|c|c|}

\hline
    Error site & Generated errors \\
    \hline
    1 & $D_{1e}(0)D_{4e}(1)$\\
    \hline
    2 & $V(D_{2o}(1)M_{2o}(1))V(D_{3o}(1)M_{3o}(1))D_{4e}(1)$\\
    \hline
    3 & $V(D_{3o}(1)M_{3o}(1))D_{4e}(1)$\\
    \hline
    4 & $D_{4e}(1)$\\
    \hline
    5 & $D_{4e}(1)$\\
    \hline
    6 & $D_{4e}(1)$\\
    \hline
    7 & $D_{4e}(1)M_{2z}(2)$\\
    \hline
    8 & $D_{4e}(2)M_{3z}(2)$\\
    \hline
    9 & $D_{4e}(2)$\\
    \hline
    10 & $\emptyset$\\
    \hline
\end{tabular}
\label{tabA1}
\captionof{table}{Generated errors for ten possible error sites for data qubits in even lines and X ancilla qubits: Each term represents 50\% X error on that data qubit/ancilla qubit, the time that the error happens is labeled by bracket (0 means initialization error on data qubit, 1 means measurement error of the first round or data qubit error after the first round of measurement and 2 has similar meaning with 1). V represents the 50\% vertical hook X error. The vertical hook error comes from the fact that the final CNOT gate copies an X error from the data qubit in odd lines to both the data qubit and the related ancilla qubit. If we consider the circuit with the feed-forward gate, error sites 5 and 10 are not accounted for.}

\end{minipage}

\begin{minipage}[c]{\textwidth}
\centering
\renewcommand\arraystretch{1.5}
\renewcommand{\thetable}{A2}
\begin{tabular}{|c|c|}

\hline
    Error site & Generated errors \\
    \hline
    1 & $M_{4z}(1)M_{4z}(2)M_{1z}(2)(D_{o}(1))D_{o}(2)$\\
    \hline
    2 & $M_{4z}(1)M_{4z}(2)M_{1z}(2)(D_{o}(1))D_{o}(2)$\\
    \hline
    3 & $M_{4z}(1)M_{4z}(2)M_{1z}(2)(D_{o}(1))D_{o}(2)$\\
    \hline
    4 & $M_{4z}(1)M_{4z}(2)M_{1z}(2)(D_{o}(1))D_{o}(2)$\\
    \hline
    5 & $M_{4z}(2)M_{1z}(2)(D_{o}(1))D_{o}(2)$\\
    \hline
    6 & $M_{4z}(2)M_{1z}(2)D_{o}(2)$\\
    \hline
    7 & $M_{4z}(2)D_{o}(2)$\\
    \hline
    8 & $M_{4z}(2)D_{o}(2)$\\
    \hline
    9 & $M_{4z}(2)D_{o}(2)$\\
    \hline
    10 & $M_{4z}(2)D_{o}(2)$\\
    \hline
\end{tabular}
\label{tabA2}
\captionof{table}{Generated errors for ten possible error sites for data qubits in odd lines and Z ancilla qubits: the symbol has a similar meaning to that in Table.A1. The brackets in $(D_{o}(1))$ means that when $D_{o}(2)M_{4z}(2)M_{1z}(2)$ exists, $D_{o}(1)$ makes no difference so it is not necessary to add it. However, in the last round that $D_{o}(2)M_{4z}(2)M_{1z}(2)$ does not exist, $D_{o}(1)$ is required. $M_{4z}(2)$ always exists because it is a detection of Rydberg decay error. Such an erasure error is regarded as a 50\% X error.}
\end{minipage}\\

There are two kinds of errors that need additional attention, both from possible error sites for data qubits in odd lines and Z ancilla qubits. First is single leakage error in error sites 1,2,3,4, which introduces successive measurement error $M_{4z}(1)M_{4z}(2)$. Such time-correlated error is common when dealing with leakage error \cite{brown2020critical,PhysRevA.100.032325}, and it does not bother when we implement a logical memory, the condition discussed in our main text \cite{timelike}. Such time-correlated errors require more syndrome extraction rounds in lattice surgery \cite{timelike}, but its effect is not clear when regarding transversal gate and correlated decoding, a more practical scenario for neutral atoms \cite{Baranes_2026,PhysRevLett.133.240602}. We leave a concrete study on it for future work.

Another important error comes from the specific dynamics of neutral atoms. If one atom decays to lower levels during a two-qubit gate, another atom is driven to the Rydberg state with $\sim O(1)$ probability, resulting in an atom loss error after the gate. See Supplementary Information S4 in \cite{wu2022erasure} for details. This is a worse condition compared to the independent leakage assumption because the data qubit and ancilla qubit leak simultaneously with $\sim O(p_e)$, instead of a second-order small term $\sim O(p_e^2)$. We refer to it as \textit{correlated leakage}. If a \textit{correlated leakage} happens in the first CNOT gate of Z syndrome measurement, leakage in the ancilla qubit introduce a data qubit error $D_{(4)o}$ after this round (see error site 1 in Table.A2) and leakage in the data qubit $D_{1o}$ introduce a data qubit error $D_{1o}$ after this round. These two errors lie in the same X logical operator ($X_2$ in Fig.\ref{figS2}), so the distance is degraded. This suggests that when considering the logical error rate of one logical operator, the logical error rate of $X_2$ should be higher than that of $X_1$ when the physical error rate $p$ is small. It can be easily verified that the degradation with distance can not be eliminated by changing the gate sequence.

\section{Two-qubit depolarization in SWAP-LRC}\label{appendix3}

\begin{figure*}[h]
\renewcommand{\thefigure}{S5}

    \centering
    \includegraphics[width=\textwidth]{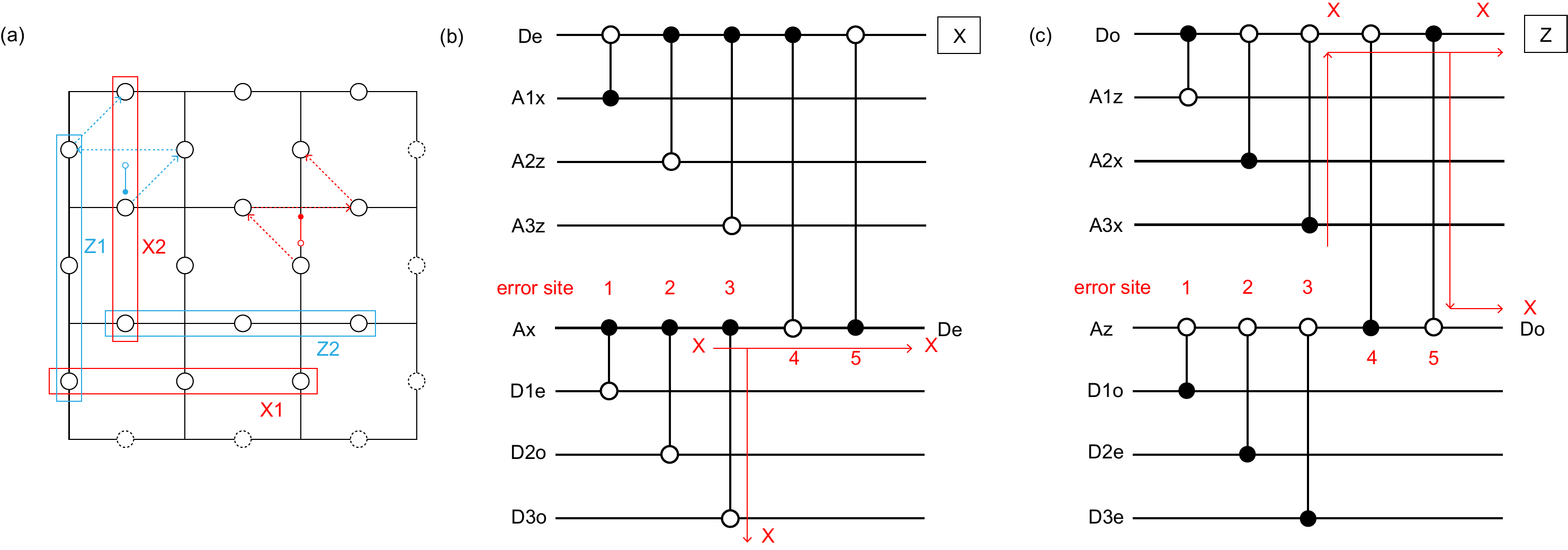}
    \captionsetup{justification = raggedright,singlelinecheck = false}
    \caption{A schematic diagram for the propagation of the two-qubit depolarization error (a) a d = 3 toric code, (b) the gate sequence and error sites related to X syndrome, (c) the gate sequence and error sites related to Z syndrome. The propagation of X-I type error in the second error site related to X syndrome is marked, and other propagation is derived similarly.}
    \label{figS5}
\end{figure*}

In this appendix, we explain how we add the Pauli error into our simulation. The Pauli error is modeled by the two-qubit depolarization $\frac{p_d}{15}\{I,X,Y,Z\}^{\otimes 2}\setminus{\{I\otimes I\}}$. For clarity, we only consider X errors and Z syndrome measurement errors. Z errors and X syndrome measurement errors can be considered in the same way. Here, we classify the Pauli error into three types, depending on whether the Pauli operator has an X operator in the ancilla qubit or data qubit. If it does not have an X operator in both qubits, it acts trivially. Namely we have $\{X_a,Y_a\}\otimes\{X_d,Y_d\}$, $\{X_a,Y_a\}\otimes\{I_d,Z_d\}$, $\{I_a,Z_a\}\otimes\{X_d,Y_d\}$ as X-X type, X-I type and I-X type. Each type has probability $\frac{4p_d}{15}$ for each CNOT gate. Then we analyse all possible gate errors independently, for both CNOT gates acting on X syndrome qubits and Z syndrome qubits. The results are listed in Table.A3 and Table.A4, and a schematic diagram is shown in Fig.\ref{figS5}. All the errors are added to the detector error model with stim \cite{Gidney_2021}.

\begin{minipage}[c]{\textwidth}
\centering
\renewcommand\arraystretch{1.5}
\renewcommand{\thetable}{A3}
\begin{tabular}{|c|c|c|c|c|c|}

\hline
    Type/Site & 1 & 2 & 3 & 4 & 5 \\
    \hline
    X-X type & $I$ & $D_{1e}(-1)$ & $D_{3o}M_3D_{4e}$ & $D_{4e}$ & $D_{4e}$\\
    \hline
    X-I type & $D_{1e}(-1)$ & $D_{3o}M_3D_{4e}$ & $D_{4e}$ & $D_{4e}$ & $D_{4e}$\\
    \hline
    I-X type & $D_{1e}(-1)$ & $D_{2o}M_2$ & $D_{3o}M_3$ & $I$ & $I$\\
    \hline
\end{tabular}
\label{tabA3}
\captionof{table}{Possible errors related to X syndrome qubits, induced by the two-qubit depolarization error from CNOT gates: D represents the data error and M represents the Z syndrome error. The subscript in D represents the time sequence that the data qubit interacts with the syndrome and the type of data qubit (in even lines or odd lines). Z syndrome measurement is always related to some data qubits in odd lines, so the subscript is the related data qubit and its time sequence. We assume the data error happens after the syndrome measurement without additional notation, but $(-1)$ represents the error that happens before the syndrome measurement. Each term represents an X error instead of a 50\% X error in the discussion in Appendix.\ref{appendix2}.}
\end{minipage}

\begin{minipage}[c]{\textwidth}
\centering
\renewcommand\arraystretch{1.5}
\renewcommand{\thetable}{A4}
\begin{tabular}{|c|c|c|c|c|c|}

\hline
    Type/Site & 1 & 2 & 3 & 4 & 5 \\
    \hline
    X-X type & $D_{1o}(-1)$ & $D_{2e}(-1)$ & $D_{3e}M_4$ & $M_4$ & $D_{4o}M_4$\\
    \hline
    X-I type & $M_4$ & $M_4$ & $M_4$ & $D_{4o}$ & $D_{4o}$\\
    \hline
    I-X type & $M_1 D_{1o}$ & $D_{2e}(-1)M_4$ & $D_{3e}$ & $D_{4o}M_4$ & $M_4$\\
    \hline
\end{tabular}
\label{tabA4}
\captionof{table}{Possible errors related to Z syndrome qubits: The setting is similar to that in Tab. A3.}
\end{minipage}

\noindent
\newpage
\bibliographystyle{unsrt}
\bibliography{ref.bib}
\clearpage

\end{document}